\documentclass[twocolumn,pdftex,letter]{jpsj3}
\usepackage{graphicx}
\usepackage{txfonts,bm,revsymb}
\usepackage{bm}
\usepackage{color}
\usepackage{epstopdf}

\def\Vec#1{\bm{#1}}

\title{
First-principle study of antimony doping effects on the iron-based superconductor CaFe(Sb$_{x}$As$_{1-x}$)$_{2}$}

\author{Yuki Nagai$^{1}$, Hiroki Nakamura$^{1}$, Masahiko Machida$^{1}$, and Kazuhiko Kuroki$^{2}$}
\inst{
\address{$^{1}$CCSE, Japan  Atomic Energy Agency, Kashiwa, Chiba 277-0871, Japan} \\
$^{2}$Department of Physics, Osaka University, Toyonaka, Osaka 560-0043, Japan
}

\abst{
We study antimony doping effects in the iron-based superconductor CaFe(Sb$_{x}$As$_{1-x}$)$_{2}$ 
by using the first-principle calculation. 
The calculations reveal that the substitution of the doped antimony atom into As of the chainlike As layers is more stable than that in FeAs layers. 
This prediction can be checked by experiments. 
Our results suggest that doping homologous elements into the chainlike As layers existing only in novel 112 system is responsible for rising up the critical temperature.  
We discuss antimony doping effects on the electronic structure. 
It is found that the calculated band structures with and without the antimony doping are similar to each other within our framework. 
}

\begin{document}
\maketitle

The discovery of the iron-based superconductor LaFeAsO$_{1-x}$F$_{x}$ has attracted much attention because of the 
high superconducting transition temperature\cite{Kamihara}. 
Many different families of iron-based materials have been found, such as $RE$FeAsO (1111 type)\cite{Kamihara}, $AE$Fe$_{2}$As$_{2}$ (122 type)\cite{Rotter}, $A$FeAs (111 type)\cite{Tapp}, and FeSe (11 type)\cite{Hsu} and so on, which contain rare-each ($RE$), alkali-earth ($AE$), and alkali ($A$) elements. 
The maximum superconducting transition temperature $T_{\rm c}$ is $55$K and all the iron-based superconductors 
consist of FeAs (FeSe) layers and various kinds of blocking layers. 

Recently, the discovery of the novel 112-type iron arsenides of Ca$_{1-x}$La$_{x}$FeAs$_{2}$, Ca$_{1-x}$Pr$_{x}$FeAs$_{2}$ and Ca$_{1-x}$$Re$$_{x}$FeAs$_{2}$ ($RE = $ Ce, Nd, Sm, Eu, and Gd) have received considerable attention\cite{Katayama, Yakita, Sala}. 
The crystal structure of the 112-type iron arsenides is monoclinic 
and consists of alternately stacked FeAs and arsenic zigzag bond layers. 
Very recently, it was reported that the simultaneous doping of isovalent P or Sb drastically improves the superconductivity in the La-doped 112 phase\cite{Kudo}. 
In addition, a large amount of Sb doping leads to a further increase in $T_{\rm c}$ to 47 K\cite{Kudo2}. 
Kubo {\it et al.} found that the enhancement of superconductivity originates from the enlargement in the lattice parameter 
$b$, which increases the As-Fe-As bond angle to be closer to the ideal tetrahedron value\cite{Lee}. 
We note that the $T_{\rm c}$ increase by the antimony substitution does not occur in any other iron pnictides. 
In order to understand the mechanism of the enhancement of $T_{\rm c}$, we need to know where the antimony is substituted, since there are two possible Sb substitution points; the chainlike As layer and the FeAs layer.

The electronic structure of CaFeAs$_{2}$ was investigated by two groups using the first-principle calculations\cite{Katayama,Wu}. 
Katayama {\it et al.} reported, with the use of the experimental lattice parameters, 
that the Fermi surfaces consist of cylindrical ones and the overall characteristics are similar to those of the high-temperature iron-based superconductors LaFeAsO. 
The band calculation by Wu {\it et al.} with the optimized lattice parameters suggested that 
the Fermi surfaces are very similar to the normal Fermi surfaces in other iron-based superconductors except for the additional 3D hole pocket and four electron cones.

In this paper, we perform first-principle calculations to investigate antimony doping effects on 
CaFe(Sb$_{x}$As$_{1-x}$)$_{2}$. 
Consequently, we find that the antimony tends to be substituted into the chainlike As layers, not the FeAs layers, since the energy of the Sb substitution into the chainlike As layers is 0.63eV per $2 \times 2 \times 1$ supercell of CaFeAs$_{2}$ lower than that into the FeAs layer.
In these calculations, we optimize the lattice parameters and atom positions.
In the case of the antimony substitution into the chainlike As layer, we 
observe an increase in the lattice parameter $b$. 
This result is consistent with the experiments. 
We also show the difference of the band dispersions with and without the Sb substitution. 

\begin{figure}[t]
  \begin{center}
    \begin{tabular}{p{0.5 \columnwidth}p{0.5 \columnwidth}}
      \resizebox{0.4 \columnwidth}{!}{\includegraphics{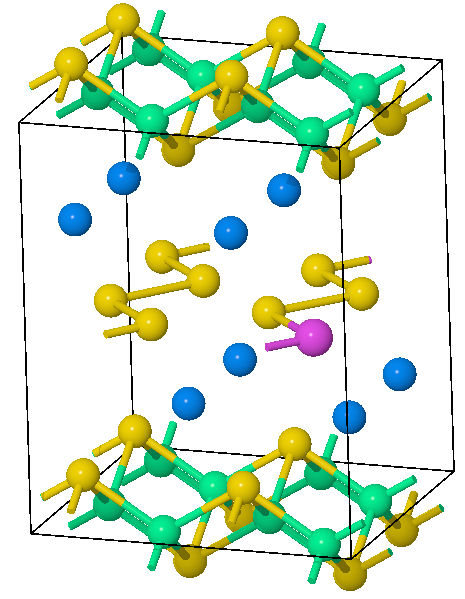}} &
      \resizebox{0.4 \columnwidth}{!}{\includegraphics{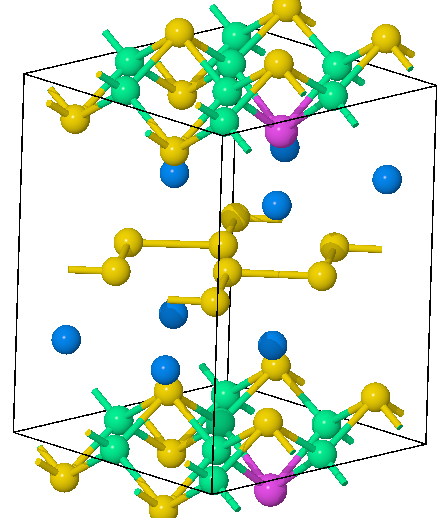}} 
    \end{tabular}
\caption{\label{fig:1}
(Color online)$2 \times 2 \times 1$ supercell of CaFeAs$_{2}$ with Sb substitution in (a) the chainlike As layer and (b) the FeAs layer
}
  \end{center}
\end{figure}

Here, we briefly show the present calculation techniques. 
The calculation package employed throughout this paper is VASP\cite{VASP}, which adopts PAW method\cite{PAW} and 
GGA exchange-correlation energy\cite{GGA}. 
In order to examine where the antimony is substituted, we prepare $2 \times 2 \times 1$ supercell of Ca$_{8}$Fe$_{8}$SbAs$_{15}$ as shown in Fig.~\ref{fig:1}. 
The chemical formula is given by CaFe(Sb$_{x}$As$_{1-x}$)$_{2}$, in which $x = 0.0625$. 
We optimize the crystal structure of CaFe(Sb$_{x}$As$_{1-x}$)$_{2}$ until the forces on all atoms become less than 0.01 eV/\AA. 
We relax atom positions, unit cell shape, and unit cell volume. 
The parameter set in the DFT calculations are given as follows. 
$\Vec{k}$-points are taken as $10 \times 11 \times 8$, and the self-consistent loops are repeated until the energy deviation becomes less than $10^{-7}$eV with the cut-off energy being 600 eV.

\begin{table*}[th]
\caption{Calculated crystal structure of CaFe(Sb$_{0.0625}$As$_{0.9375}$)$_{2}$ with and without the antimony substitution. ``$x$ \%'' denotes that the length becomes large by $x$ percent from the experimental value. }
\begin{center}
\begin{tabular}{ccccccc}
\hline \hline
 & a (\AA) & b (\AA) & c (\AA) & $\alpha$ ($^{\circ}$) & $\beta$ ($^{\circ}$) & $\gamma$ ($^{\circ}$) \\
\hline \hline Experimental structure\cite{Katayama} 
 & 3.94710(10)
 & 3.87240(10)
 & 10.3210(7)
  & 90 
  & 91.415(6)
  &90
\\
 Optimized structure
 & 3.955536 (+0.21 \%)
  & 3.893495 (+0.54 \%)
   & 10.091231 (-2.23 \%)
   & 90.000000
   & 91.184438
   & 90.000000
\\
 Sb in chainlike As layer
 &
 3.978106 (+0.786 \%) 
  &
 3.9266745 (+1.40 \%)
   & 
10.088251 (-2.26 \%)
   & 90.001299
   & 91.067423
   & 89.998870
\\
 Sb in FeAs layer
 &
 3.9837315 (+0.93 \%)
  & 
 3.9137245 (+1.07 \%)
  & 
10.120249  (-1.95 \%)
   & 90.000294
   & 91.257985
   & 90.002559
\\
\hline \hline
\end{tabular}
\end{center}
\label{table:1}
\end{table*}%

Now, let us discuss the calculated crystal structure of CaFe(Sb$_{0.0625}$As$_{0.9375}$)$_{2}$. 
The obtained lattice parameters are summarized in Table I. 
The parameters in the experimental structure for Ca$_{1-x}$La$_{x}$FeAs$_{2}$ was reported in Ref.~\citen{Katayama}. 
It should be noted that the lattice parameter $c$ in the optimized crystal structure is smaller than that obtained by experiments, which is usually found in iron-based materials. 
This might be originated from the fact that the Van der Waals force can not be accurately taken into consideration by the DFT calculation in the layer-type materials. 
We find that 
the antimony tends to be substituted into the chainlike As layer, since the calculated energy of the antimony substitution into the chainlike As layers is 0.63eV per $2 \times 2 \times 1$ supercell of CaFeAs$_{2}$ lower than that into the FeAs layer. 
Focusing on the crystal structure with the antimony substitution into the chainlike As layer, 
the lattice parameters $a$ and $b$ increases with the antimony doping. 
The calculation suggests that the unit volume also increases with the antimony doping, 
since the value of the $c$ axis in the antimony doped system is similar to that in the optimized system. 
The calculations reveal that the substitution of the doped antimony atom into the chainlike As layers is more stable than that in FeAs layers. 
This prediction can be checked by experiments. 
Our results suggest that the chainlike As layers, existing only in novel 112 system, have crucial role for increasing the critical temperature.

Phenomenologically, it is known that the $T_{\rm c}$ of the iron-based superconductors is optimized around the anion height of 1.38 \AA, 
or the As-Fe-As bond angle of 109$^{\circ}$\cite{Lee,Mizuguchi:2011es}. 
From this viewpoint, $T_{\rm c}$ increase may be expected more when antimony substitution occurs in the chainlike As layer, because 
in this case the height decreases from 1.418 \AA \:
toward 1.38 \AA, and the angle increases from 108.079$^{\circ}$ toward 109$^{\circ}$. 
However, there are exceptions for these phenomenological rules of optimizing $T_{\rm c}$, and it is not certain whether the rules apply to the present material. 
Therefore, we need to see how these changes in the lattice parameters affect the electronic structure in the following.

\begin{table}[t]
\caption{Anion height from Fe layer and As-Fe-As bond angle in calculated crystal structures of CaFe(Sb$_{0.0625}$As$_{0.9375}$)$_{2}$ with the antimony substitution. ``$x$ \%'' denotes that the length becomes large by $x$ percent from the experimental value. }
\begin{center}
\begin{tabular}{ccc}
\hline \hline
 & Anion height  (\AA) & Mean As-Fe-As angle $\alpha$ ($^{\circ}$)  \\
\hline Experimental structure\cite{Katayama} 
 & 1.418149
 & 108.079
\\
 Optimized structure
 & 1.26538 (-10.77 \%)
  & 114.36483 (+5.82 \%)
\\
 Sb in chainlike As layer
 &
1.24972 (-11.88 \%) 
  &
115.380 (+6.76 \%)
\\
 Sb in FeAs layer
 &
 1.283721 (-9.48 \%)
  & 
 113.9733 (+5.45 \%)
\\
\hline \hline
\end{tabular}
\end{center}
\label{table:2}
\end{table}%

\begin{figure}[t]
  \begin{center}
    \begin{tabular}{p{1 \columnwidth}}
      \resizebox{1 \columnwidth}{!}{\includegraphics{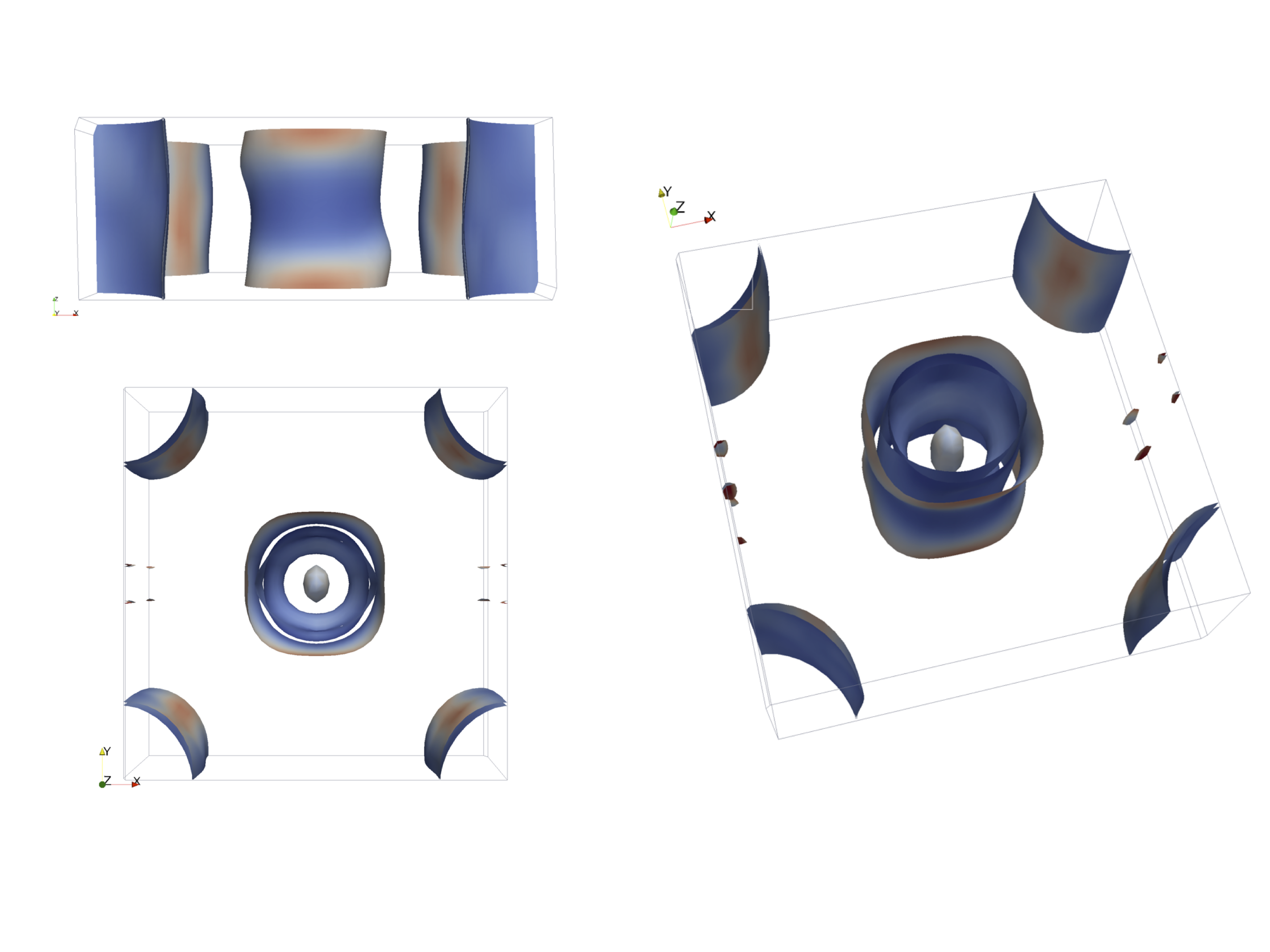}} 
    \end{tabular}
\caption{\label{fig:4}
(Color online) Fermi surfaces for the parent compound CaFeAs$_{2}$.
}
  \end{center}
\end{figure}

We first study the Fermi surfaces for the parent compound CaFeAs$_{2}$.  
We use the structural parameters obtained by the experiment in Ref.~\citen{Katayama}. 
The calculation results show that there are three two-dimensional cylindrical Fermi surfaces and one three-dimensional ellipsoidal Fermi surface around ${\rm \Gamma}$ point, two cylindrical Fermi surfaces around ${\rm A} = (1/2,1/2,0)$ point, and one very narrow cylindrical Fermi surfaces near ${\rm B} = (1/2,0,0)$ point, as shown in Fig.~\ref{fig:4}.  
The Fermi surfaces are similar to those in 1111-type iron-based superconductors, except for the three-dimensional ellipsoidal and very narrow cylindrical Fermi surfaces. 
Note that Wu {\it et al.} showed the Fermi surfaces with the optimized lattice parameters in Ref.~\citen{Wu}. 
The difference from the Fermi surfaces shown in Ref.~\citen{Wu} is 
the position of the three-dimensional ellipsoidal Fermi surface. 
In their calculation, there is the ellipsoidal Fermi surface around ${\rm Z} =(0,0,1/2)$ point.

\begin{figure}[t]
  \begin{center}
    \begin{tabular}{p{\columnwidth}}
    	\begin{center}
      \resizebox{ 1\columnwidth}{!}{\includegraphics{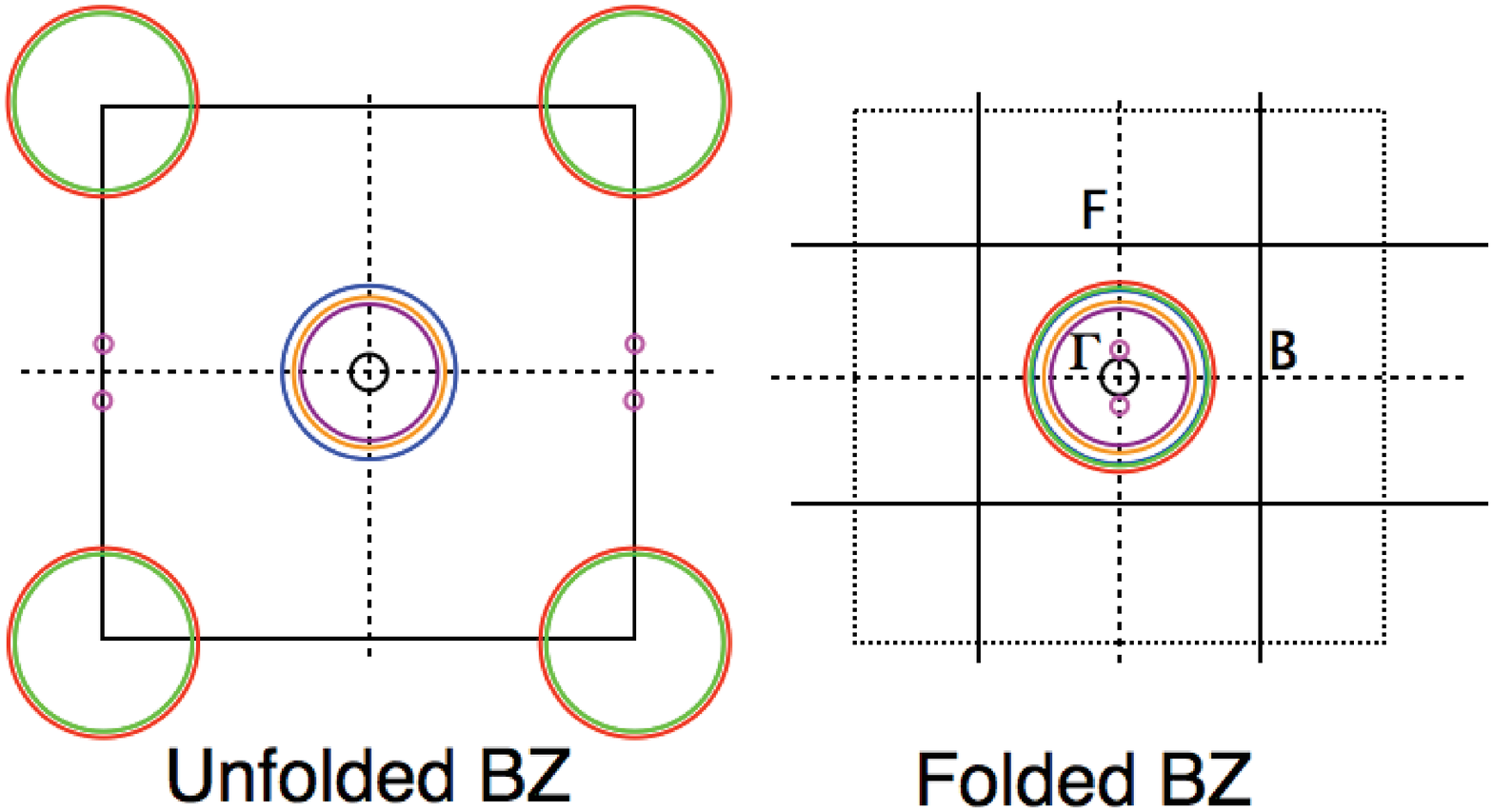}} 
      \end{center}
    \end{tabular}
\caption{\label{fig:fold}
(Color online) Schematic figure of the unfolded and folded Brillouin zones at $k_{z} = 0$. 
}
  \end{center}
\end{figure}

\begin{figure}[t]
  \begin{center}
    \begin{tabular}{p{\columnwidth}}
    	\begin{center}
      \resizebox{ 1\columnwidth}{!}{\includegraphics{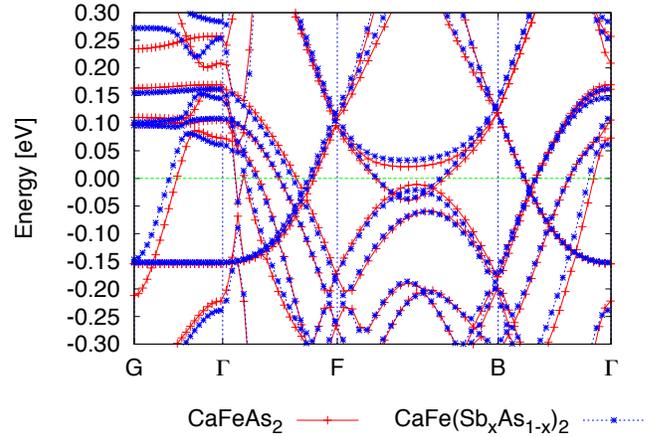}} 
      \end{center}
    \end{tabular}
\caption{\label{fig:2}
(Color online) Electronic structures with and without the antimony substitution onto the chainlike As layer. 
The experimental structural parameters are used for the parent compound and the modified ones are used for the doped compound. 
}
  \end{center}
\end{figure}

Next, let us discuss antimony doping effects on the electronic structure. 
To calculate the band structure of the parent compound CaFeAs$_{2}$, we use the structural parameters obtained by the experiment in Ref.~\citen{Katayama}. 
Note that we do not use the optimized structure, since the lattice parameters obtained by the optimization are much differ from those obtained by the experiment.
Thus, the optimization technique does not allow to evaluate accurately the electronic structure. 
To assess the electronic structure of the antimony doped compound, 
we modify the the structural parameters of the parent compound obtained by the experiment. 
With fixing the ratios of $a_{0}$/$a_{\rm Sb}$, $b_{0}$/$b_{\rm Sb}$, $c_{0}$/$c_{\rm Sb}$, and $h_{0}$/$h_{\rm Sb}$, 
we adopt the lattice parameters of the $2 \times 2 \times 1$ supercell of CaFeAs$_{2}$ are $a = 7.939243$ \AA, $b = 7.810799$ $\AA$ and $c = 10.317950$ \AA. 
Here, $a_{0(\rm Sb)}$, $b_{0(\rm Sb)}$ and $c_{0(\rm Sb)}$ are the lengths of the unit cell, and $h_{0({\rm Sb})}$ is the anion height from Fe layer in parent (Sb-doped) compound, respectively. 
This treatment with fixing ratios of the lattice parameters and the anion height was used in the calculation to discuss the pressure effect in other pnictides\cite{SuzukiKurokiun}. 
The band structures are obtained in folded Brillouin zone of the supercell.
We show the unfolded and folded Brillouin zones at $k_{z} = 0$ in Fig.~\ref{fig:fold}. 
The symmetric points in the triclinic Brillouin zone are ${\rm \Gamma} = (0,0,0)$, ${\rm B} = (1/2,0,0)$, ${\rm F} = (0,1/2,0)$ and ${\rm G} = (0,0,1/2)$. 
In this Brillouin zone, there are the very-narrow cylindrical Fermi surfaces on the ${\rm \Gamma}$-F line. 
In the case of the antimony substitution with the use of the above technique, the band along G-${\rm \Gamma}$-line shifts, where the size of the three-dimensional Fermi surface becomes relatively large as shown in Fig.~\ref{fig:2}. 
These results suggest that the calculated band structures with and without the antimony doping are similar to each other. 
In terms of the enhancement of  $T_{\rm c}$ with the antimony doping, 
there are three possibilities of the role of antimony doping effects as follows. 
First, there is a possibility that these small change of the band structure brings the large increase of $T_{\rm c}$. 
A theoretical microscopic calculation is required, such as solving the Eliashberg equations. 
The second possibility is that the $T_{\rm c}$ enhancement in this material is not related to the change of the band structure. 
One might have to consider the local change of the electronic states which is hard to be treated by the band calculation and/or the possibility of the increase of the electron-electron interaction due to the antimony doping. 
The third one is the possibility that 
the adopted lattice parameters with fixing the ratios of the structural parameters is not  suitable to describe the electric structure with and without the antimony doping. 
It is important to find out the appropriate structure which can describe the electric structure with the antimony doping correctly. 
It remains as a future work. 

%

\begin{figure*}[th]
\vspace{10mm}
\begin{center}
\begin{tabular}{p{ 0.48\textwidth} p{  0.48\textwidth} } 
\raisebox{5cm}{(a)} \resizebox{  0.48\textwidth}{!}{\includegraphics{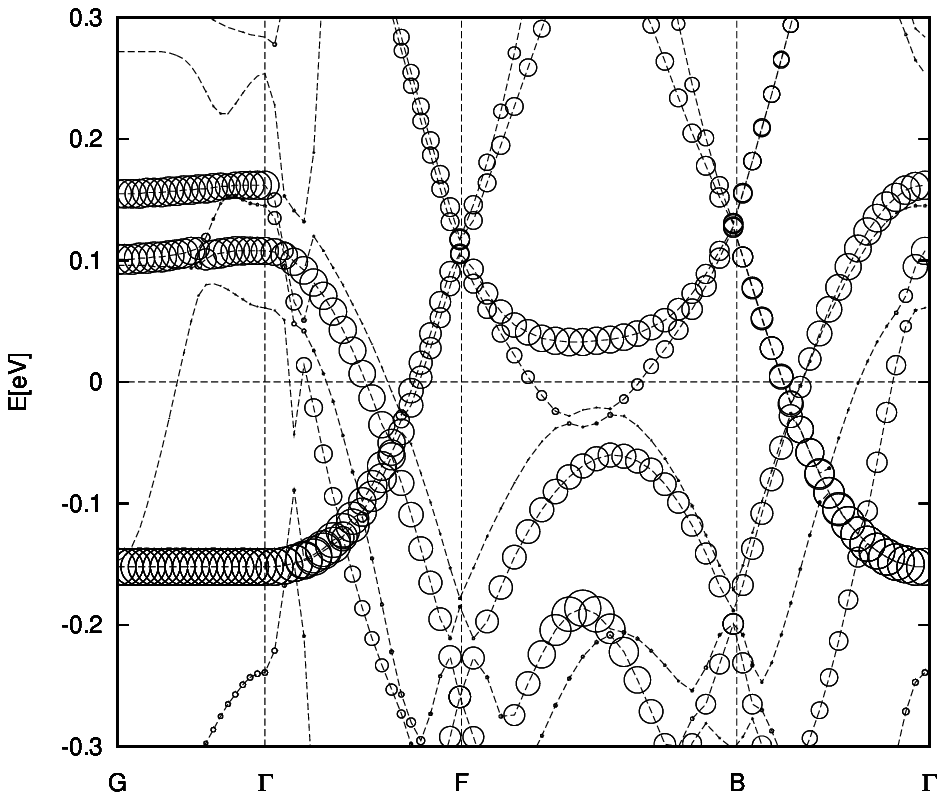}} &\raisebox{5cm}{(b)}  \resizebox{ 0.48\textwidth}{!}{\includegraphics{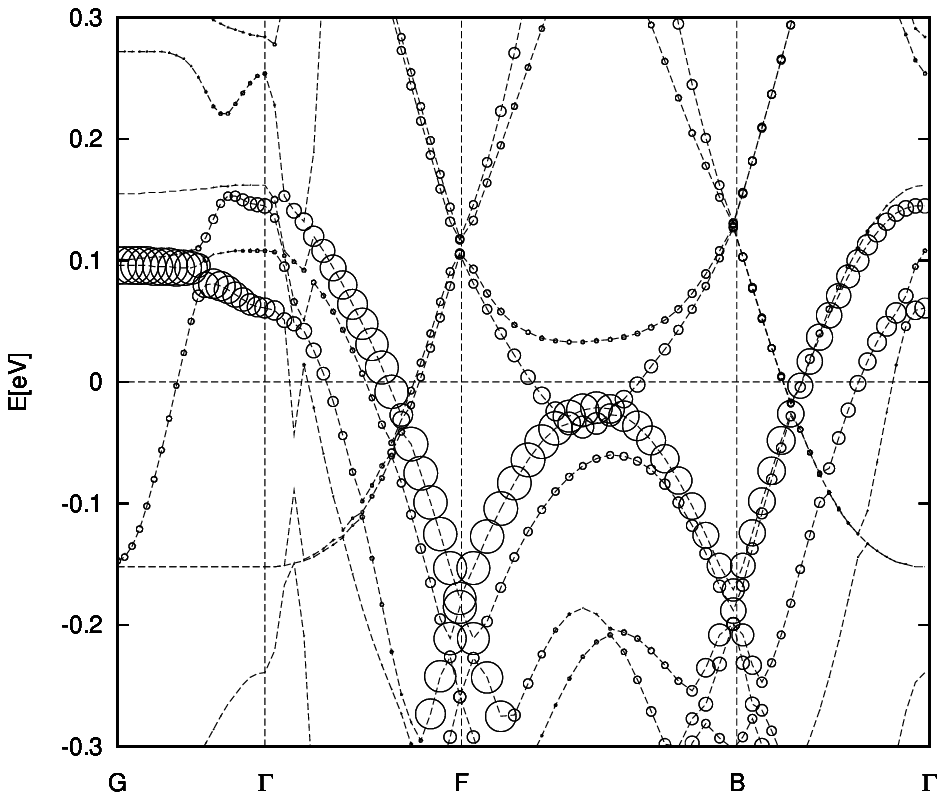}}  \\
\raisebox{5cm}{(c)}  \resizebox{  0.48\textwidth}{!}{\includegraphics{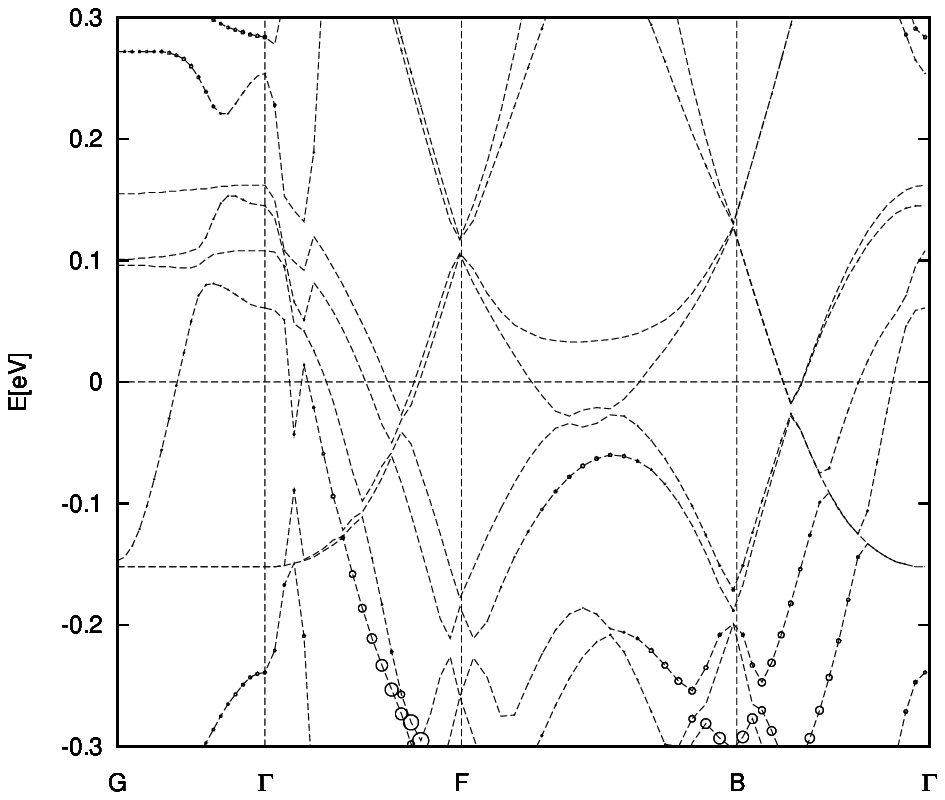}} &\raisebox{5cm}{(d)}  \resizebox{  0.48\textwidth}{!}{\includegraphics{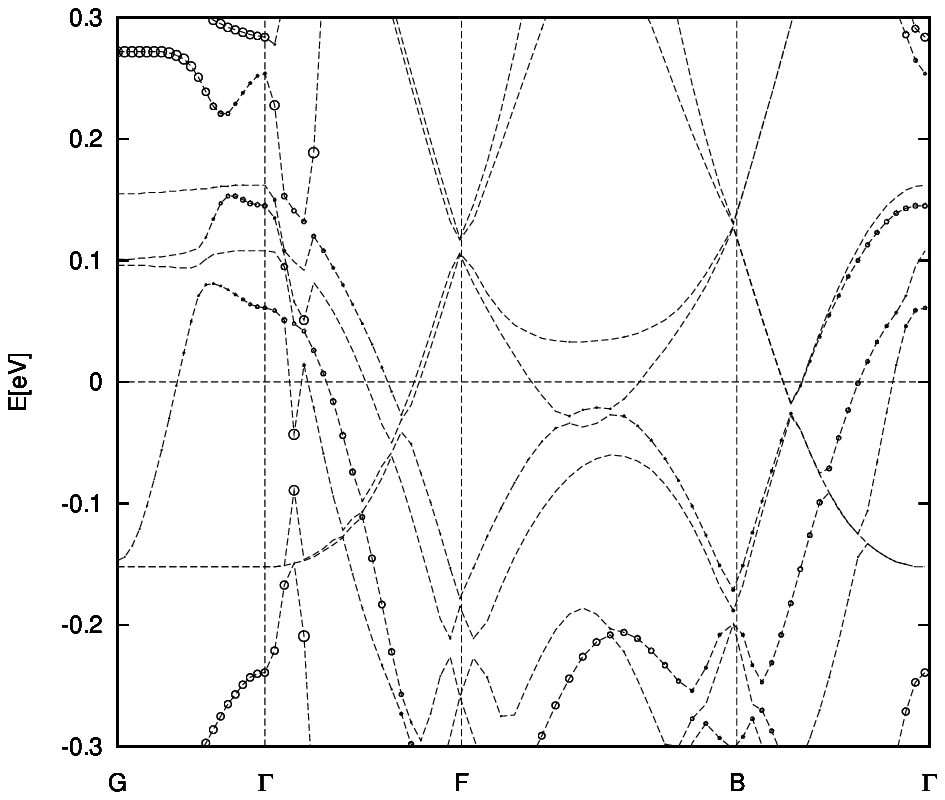}} 
\end{tabular}
\end{center}
\caption{\label{fig:orb} 
The orbital characters of the bands for CaFe(Sb$_{0.0625}$As$_{0.9375}$)$_{2}$. 
The size of the circle denotes the contribution of (a) Fe-$d_{xz,yz}$- (b) Fe-$d_{xy,x^{2}-y^{2}}$- (c) Fe-$d_{z^{2}}$- (d) chainlike-As-$p$-orbitals. 
}
\end{figure*}

Finally, let us comment on the possible role of the chainlike As layers. 
Along the ${\rm \Gamma}$-F line, there is a steep band intersecting the Fermi level, which originates from the As-$p$ orbitals of the chainlike layers as shown in Fig.~\ref{fig:orb}. 
This means that the chainlike As layers in this material are not insulating. 
This is unusual compared to other iron-based superconductors, where the superconducting transition temperature becomes higher when the ``insulating'' layer between the FeAs layers is more insulating. 
This also might be of some importance to understand the relatively high $T_{\rm c}$ of CaFeAs$_{2}$.

In summary, 
we studied the antimony doping effect in the iron-based superconductor CaFe(Sb$_{x}$As$_{1-x}$)$_{2}$ with the 
use of the first-principle calculation. 
The calculations revealed that the substitution of the doped antimony atom into the chainlike As layers is more stable than that in FeAs layers. 
This prediction can be checked by experiments. 
Our results suggested that the chainlike As layers, which exist only in novel 112 system, have crucial role for increasing the critical temperature.  
With the antimony substitution, the band along G-${\rm \Gamma}$-line shifts, where the size of the three-dimensional Fermi surface becomes large. 
Since the calculated band structures with and without the antimony doping are similar to each other, 
more theoretical efforts will be necessary to clarify the origin of the $T_{\rm c}$ increase upon antimony substitution.

\begin{acknowledgment}
The calculations have been performed using the supercomputing 
system PRIMERGY BX900 at the Japan Atomic Energy Agency. 
This study was partially supported by JSPS KAKENHI Grant Number 24340079 and 26800197. 
\end{acknowledgment}


\begin{thebibliography}{9}
\bibitem{note}
We use the average of the angles $\alpha$ in the supercell.
\end{thebibliography}
\end{document}